\documentclass[11pt]{article} 
\renewcommand{\Box}{\hspace*{\fill}\rule{2mm}{2mm} \par\medskip}

\newtheorem{theorem}{Theorem}
\newtheorem{corollary}{Corollary}

\newcommand{\junk}[1]{}

\begin{document}

\title{String Matching in ${\tilde O}(\sqrt{n}+\sqrt{m})$ Quantum Time}
\author{H. Ramesh\thanks{Indian Institute of Science, Bangalore, 
ramesh@csa.iisc.ernet.in.}
\and V. Vinay\thanks{Indian Institute of Science, Bangalore, 
vinay@csa.iisc.ernet.in.} }
\date{}
\maketitle
\begin{abstract}
We show how to determine whether a given pattern $p$ 
of length $m$ occurs in a given text $t$ of length $n$
in ${\tilde O}(\sqrt{n}+\sqrt{m})$\footnote{${\tilde O}$ allows for logarithmic factors in $m$ and $n/m$} time, with inverse polynomial
failure probability.
This algorithm combines quantum searching algorithms
with a technique from parallel string matching,
called {\em Deterministic Sampling}.


\end{abstract}

\section{Introduction}

We consider the following problem: given a text $t$ of length $n$ and
a pattern $p$ of length $m$, does $p$ occur in $t$?
This question requires $\Theta(n+m)$ time classically, using 
one of several known algorithms, e.g., the Knuth-Morris-Pratt
algorithm or the Boyer-Moore algorithm.
Note that the above problem is slightly 
different from the usual string matching problem which
requires finding all occurrences of the pattern in the text.

We explore the above question on a quantum machine.
Our starting point is the algorithm due to Grover\cite{G}
which searches for an element in an unordered database in
$O(\sqrt{n})$ time.
Boyer, Brassard, Hoyer and Tapp\cite{Bo}
gave a tighter analysis of Grover's algorithm
and also showed how to handle the case when
the number of items searched for is unknown.
D\"urr and  Hoyer\cite{Du} used Grover's algorithm to find
the minimum in $O(\sqrt{n})$ time.

Finding whether the pattern matches 
somewhere in the text is akin to searching in an unordered
database; the only issue is that checking 
one element of this database, i.e., a text position,
for an occurrence of the pattern takes $O(m)$ time.
In fact, this can be speeded up to 
$O(\sqrt{m})$ by viewing the act of checking whether 
the pattern matches at a particular text position
as the act of finding a mismatch amongst $m$ elements;
this search can be performed in $O(\sqrt{m})$ time
with constant failure probability.
This gives an overall time complexity of $O(\sqrt{nm})$
(actually, there are additional logarithmic factors,
as will be described later).

In this paper, we show how this complexity can be
improved to $O(\sqrt{n}\log \sqrt{\frac{n}{m}}\log m+\sqrt{m}\log^2 m)$,
by combining the above quantum search paradigm with 
a standard technique from parallel string matching,
called {\em Deterministic Sampling}, due to Vishkin\cite{Vi}.
Our algorithm will work with constant failure probability.
Thus, if the pattern occurs in the text, it will 
return some occurrence of the pattern
in the text (or the leftmost occurrence, if needed)
in $O(\sqrt{n}\log \sqrt{\frac{n}{m}}\log m+\sqrt{m}\log^2 m)$
time, with probability which is a constant strictly more than 1/2.
And if the pattern does not occur in the text
then the algorithm will say so with probability which is also
a constant strictly more than 1/2.
The failure probability can be decreased to inverse polynomial 
at the expense of further logarithmic factors.
Finally, note that the second component of the running time
above will be due to pattern preprocessing.

\section{Preliminaries}

We use the following theorem based on Grover's\cite{G}
database searching algorithm.
The theorem itself is due to Boyer et al.\cite{Bo}.

\begin{theorem}
\label{th1}
Given an oracle evaluating to 1 on at least $t\geq 1$ 
of the elements in an unordered database of size $n$,
there is a quantum algorithm which returns the index
of a random element on which the oracle evaluates to 1,
with probability at least 3/4, 
in $O(\sqrt{\frac{n}{t}})$ time and oracle calls.
\end{theorem}

We will also need the following theorem for finding the minimum 
element in a database, due to D\"urr and  Hoyer\cite{Du}.

\begin{theorem}
\label{th2}
Given a comparison oracle,
there is a quantum algorithm which finds the index of
the minimum element in an unordered database of size $n$,
with probability at least 3/4, 
in $O(\sqrt{n})$ time.
\end{theorem}

We will assume a basic oracle which will compare
a text and a pattern character or two pattern characters
in $O(1)$ time.
Our aim then is to use this oracle to develop suitable oracles which
will enable solving the string finding problem
in ${\tilde O}(\sqrt{n}+\sqrt{m})$ time.
These new oracles could be probabilistic, i.e., they
will give the correct answer with constant probability.
We derive the following corollary to Theorems \ref{th1} and \ref{th2},
in order to use probabilistic oracles.
First, we define a probabilistic oracle formally.

\medskip
\noindent
{\bf Probabilistic Oracles.} These are oracles which  evaluate
to 1 on {\em good} elements (i.e., those which are being searched for)
 with probability at least 3/4 and to  0 on bad elements,
with probability at least 3/4.
A {\em probabilistic comparison oracle} gives the 
correct answer with probability at least 3/4.

\begin{corollary}
\label{cor}
Given a probabilistic oracle and a database with $t\geq 1$ good elements,
there is a quantum algorithm which returns the index
of a random good element with probability at least 3/4
in $O(\sqrt{\frac{n}{t}}\log \sqrt{\frac{n}{t}})$ time and oracle calls.
Similarly, given a probabilistic comparison oracle,
there is a quantum algorithm which finds the index of
the minimum element with probability at least 3/4, 
in $O(\sqrt{n}\log \sqrt{n})$ time.
\end{corollary}
{\bf Proof.} For the searching problem,
each original oracle call in
Theorem \ref{th1}  gets replaced by 
$O(\log \sqrt{\frac{n}{t}})$ calls
and the majority result is taken.
This is to ensure that, with constant probability, 
none of the original $\sqrt{\frac{n}{t}}$ oracle calls returns a 1
on a bad element or a 0 on a good element.
Similarly, for the minimum finding problem,
each original oracle call in
Theorem \ref{th2}  gets replaced by 
$O(\log \sqrt{n})$ calls. \Box

\medskip
Our algorithm will use oracles whose running time
will not be a constant. 
Therefore, the search time will be obtained by multiplying
the time given by the above corollary with the
time taken per oracle call.

\medskip
We will require the following facts about strings.

\medskip
\noindent
{\bf Periodicity.}  A string $p$, $|p|=m$, is said to be {\em aperiodic}
if any two instances of the string, one shifted to the
right of the other by at most $m/2$, differ in some column.
A string which is not aperiodic is called {\em periodic}.
A periodic string $p$ has the form $v^ku$,  where $v$ 
cannot be expressed as a concatenation of several instances of 
a smaller string, $u$
is a prefix of $v$, and $k\geq 2$.
$|v|$ is said to be the {\em period} of $p$.

\section{ The ${\tilde O}(\sqrt{n}\sqrt{m})$ Time Algorithm}
\label{nm}

Consider using Grover's algorithm in conjunction 
with the following natural probabilistic oracle $f()$ 
to solve the string matching problem:
$f(i)=1$ if  the pattern matches 
with left endpoint aligned with text position $i$,
and $f(i)=0$ with probability at least 3/4, otherwise.
This probabilistic oracle can be implemented so that it runs in 
${\tilde O}(\sqrt{m})$ time as follows.

The idea is to implement the oracle $f(i)$ using Grover's
algorithm itself, using a deterministic oracle $g(i,j)$ which looks
for a mismatch at location $j$.
$g(i,j)=1$ if and only if the pattern with left endpoint
aligned with  $t[i]$ mismatches at location $p[j]$.
By Theorem \ref{th1}, such a location, 
if one exists, can be found in $O(\sqrt{m})$ time with  
probability at least 3/4.
We set $f(i)=0$ if  Grover's algorithm with
oracle $g(i,j)$ succeeds in finding a mismatch, and $f(i)=1$ otherwise.
Using Corollary \ref{cor}, the time taken to search using oracle $f(i)$
is $O(\sqrt{n}\sqrt{m}\log \sqrt{n})$ and the
success  probability is at least 3/4.

Next, we give the faster algorithm, first for aperiodic strings and then for
periodic strings.

\section{The ${\tilde O}(\sqrt{n}+\sqrt{m})$ Time Algorithm:  Aperiodic Patterns}
\label{aper}

The above oracle $f(i)$ was too expensive to get 
an ${\tilde O}(\sqrt{n}+\sqrt{m})$ bound. 
A faster  oracle will clearly improve the time.
We do not know how to get a faster oracle directly.
However, we reorganize the computation as described below
and then use a faster oracle followed by a slower one,
speeding up the algorithm on the whole.

We partition the text into blocks of
length $m/2$ and use Grover's algorithm to search for
a block which contains an occurrence of the 
pattern.
This is done using a probabilistic oracle $h(i)$ (to be described
later), which evaluates
to 1 with probability at least 3/4 if block $i$ has a
match of the pattern (with left endpoint starting in block $i$),
and evaluates to 0 with probability at least 3/4, otherwise.
Note that by aperiodicity, the pattern can match with left endpoint 
at  at most one text position in block $i$.
$h(i)$ will take $O(\sqrt{m}\log m)$ time.
It follows from Corollary \ref{cor} that
the time taken for searching with the oracle $h(i)$ 
will be $O(\sqrt{\frac{n}{m}}\sqrt{m}\log \sqrt{\frac{n}{m}}\log m=\sqrt{n}\log
\sqrt{\frac{n}{m}}\log m)$ and the success probability is a constant.

The oracle $h(i)$ itself will run in two steps.
The first step will use deterministic sampling and 
will takes $O(\sqrt{m}\log m)$ time with constant success probability.
This step will eliminate all but at most one of the
pattern instances with left endpoint in block $i$.
The second step will check whether this surviving instance
matches the text using the $g()$ oracle defined 
in Section \ref{nm};
this will take $O(\sqrt{m})$ time with constant success probability.
We describe the two steps next.

\subsection{Step 1: Deterministic Sampling}

The oracle is based on the following theorem 
 due to Vishkin \cite{Vi}.
 
\begin{theorem}
\label{ds}
{\bf [The  Deterministic Sampling Theorem]}
Let $p$ be aperiodic.
Consider $\frac{m}{2}$ instances of $p$,
with successive instances shifted one step to 
the right.
Let these instances be labelled from $1$ 
to $\frac{m}{2}$, from left to right.
Then there exists an instance $f$, 
and a set of at most $O(\log m)$ positions 
in $p$, called the {\em deterministic sample} 
with the following property: 
if all positions corresponding to
the deterministic sample in instance $f$ of $p$ 
match the text, then none of the other
instances of $p$ above can possibly match the text.
\end{theorem}
{\bf Proof.} 
By aperiodicity,
there is a column which contains two distinct characters
and stabs all the pattern instances;
pick any character in that column
which is not in majority.
Remove all pattern instances which do not have this
character in the column being considered.
Repeat $O(\log m)$ times until only one pattern instance
remains.
Then $f$ is the label of the instance which remains
and the columns chosen give the deterministic sample.\Box

\medskip
Assume that a deterministic sample for the pattern has been 
precomputed. 
We will describe this precomputation later.

\medskip
We now describe the first step in  $h(i)$, where $i$ is a
block number. 
We use Grover's algorithm in conjunction with
the deterministic oracle $k(i,j)$ which evaluates
to $1$ if and only if the $j$th instance of the
pattern (amongst those instances with left endpoint
in block $i$) matches the text on its deterministic 
sample.
Clearly, $k(i,j)$ takes $O(\log m)$ time.
The search using $k(i,j)$ takes 
$O(\sqrt{m}\log m)$ time by Theorem \ref{th1}.
This search returns an instance $j$ of the
pattern  with left endpoint in block $i$
which has its deterministic sample matching the text
(if such an instance exists), with probability at least 3/4.

\subsection{Step 2: Direct Verification}

Next, we use another application of Grover's 
search to determine whether or not instance
$j$ determined above in block $i$ matches 
the text; this is done as in Section \ref{nm}
using the deterministic oracle $g()$.
It succeeds with probability 3/4, and 
takes time $O(\sqrt{m})$.

\medskip
Thus, $h(i)$ returns a 1 with probability at least 3/4 
if block $i$ contains a match of the pattern, and 0 
with probability at least 3/4, otherwise.
The time taken to search using $h(i)$ is
$O(\sqrt{n}\log\sqrt{\frac{n}{m}}\log m)$, as claimed above, 
 and the success probability is at least 3/4.

Once a block $i$ is found in which $h(i)$ evaluates to
1, a search using oracle $k(i,j)$ on block $i$
gives the unique pattern instance $j$ 
with left endpoint in $i$ whose deterministic sample
matches; the success probability is at least 3/4. 
Another search using the oracle $g()$ determines 
if this pattern instance mismatches the text;
the success probability (in finding a mismatch, if any) is again at least 3/4.
Thus, the total time taken is 
$O(\sqrt{n}\log\sqrt{\frac{n}{m}}\log m)$, 
and the probability of success is as follows.

If the pattern occurs in the text, then 
with probability 3/4, the
search with $h()$ will return a block containing 
a match of the pattern;
subsequently, with probability 3/4, the search with $k(i,j)$ will
return a matching pattern instance, and the last search with $g()$ will
not discover a mismatch with probability 1.
Thus an occurrence of the pattern in the text will
be found with $(3/4)^2> 1/2$ probability, as
claimed.
And if the pattern does not occur in the text, 
then the last search with $g()$ will determine
a mismatch with probability at least $3/4>1/2$,
as required.

Finally, note that  the leftmost occurrence of the pattern can be determined
using the minimum finding algorithm in Corollary \ref{cor} 
to first find the leftmost block with $h(i)$ evaluating to
1, and subsequently, searching within that block as above.
The time taken and the success probability are 
as in the previous paragraphs.

\section{Pattern Preprocessing for Aperiodic Patterns}

We show how to determine the deterministic sample
in $O(\sqrt{m}\log^2 m)$ time.

\medskip
\noindent
{\bf Determining the Deterministic Sample.}
Imagine $m/2$ copies of the pattern placed as
in Theorem \ref{ds}. 
Determining the sample will proceed in $O(\log m)$ stages.
In each stage, some column and a character in that
column will be identified; all surviving pattern
copies which do not have this character in this column
will be eliminated from future stages.
This will continue until only one pattern copy
remains uneliminated.
Each stage will take $O(\sqrt{m}\log m)$ time and
will have a constant success probability (where 
we count a success if the surviving pattern copies
halve in cardinality).

A stage proceeds as follows.
First, a column containing two distinct characters
amongst the surviving pattern copies is found,
with  constant probability.
This column will also have the property that 
all surviving pattern copies are stabbed by it.
Two distinct characters  in the above column are also
identified.
One of these two characters is chosen at random
as the next character in the sample.
Clearly, the number of stages is $O(\log m)$ with
inverse polynomial (in $m$) failure probability
because the probability of the number of surviving
pattern instances halving in a stage is at least a constant.

It remains to describe how a column containing two distinct characters
amongst the surviving pattern copies is found,
with constant probability.
Before describing this we need to mention how to find
the leftmost and rightmost surviving patterns in a stage.
This is done using the minimum finding 
algorithm of D\"{u}rr and Hoyer in conjunction
with an oracle which indicates
which pattern copies are consistent with the
already chosen deterministic sample points;
this oracle takes $O(\log m)$ time per call,
giving an $O(\sqrt{m}\log m)$ time algorithm for
finding the leftmost/rightmost  surviving pattern copy, 
by Theorem  \ref{th2}.
The success probability is at least 3/4.
Now, we can describe the algorithm for finding
a column with two distinct characters.

First, the leftmost and rightmost surviving
pattern copies are found as above.
Then a column in which these two copies differ is 
found  using Grover's algorithm in conjunction
with a suitable oracle in $O(\sqrt{m})$ time;
this step succeeds with probability at least 3/4.
Given a column,  this oracle determines whether or not
the two pattern copies above differ in this column.
By Theorem \ref{th1}, searching for a column with
two distinct characters using this oracle takes 
$O(\sqrt{m})$ time and succeeds with probability 3/4.

Thus, in time $O(\sqrt{m}\log m)$,
a column containing two distinct characters and 
stabbing all surviving pattern copies is found,
with constant probability; it is easily seen
that two distinct characters in this column
are also found in this process.

The total time taken in determining the
deterministic sample is thus $O(\sqrt{m}\log^2 m)$.

\section{Handling Periodic Patterns}

We sketch briefly the changes required to the above algorithm 
in order to handle periodic pattern.

For periodic patterns, the above preprocessing algorithm will not 
terminate with a single pattern copy but rather with 
several copies shifted $|v|$ steps
to the right successively.
When a stage is reached when the only surviving 
copies are the periodically shifted copies above,
then the search for a heterogeneous column
in the next  $\Theta(\log m)$ stages will fail.
Note that for aperiodic patterns this 
behaviour happens with low, i.e., inverse
polynomial probability.

At this point, we determine the period $|v|$
using two instances of the minimum finding algorithm.
The first instance finds the leftmost surviving 
copy and the second the second leftmost;
the difference of their offsets is the period.
This takes $O(\sqrt{m}\log m)$ time,
using the oracle which checks for consistency with
the deterministic sample and also compares offsets.
Given the period $|v|$, the following changes now need
to be made to the text processing part.

Recall the oracle $h(i)$ from Section \ref{aper};
this oracle determines whether there is a pattern
instance with left endpoint in block $i$ which 
matches, first on its deterministic sample,
and then on the whole.
This oracle is modified as follows.

$h(i)$ will first determine the leftmost and
the rightmost pattern instances with left endpoints
in $i$ which match on their respective deterministic
sample points; this is done using the
minimum finding algorithm and takes $O(\sqrt{m}\log m)$ time
with success probability at least 3/4 (see Theorem \ref{th2},
this success probability can be made arbitrarily close to 1
by repeating).
Let these two instance have left endpoints
at text positions $k$ and $l$ respectively.

Next, $h(i)$ finds the longest substring (with length at most $m$)
starting at the right boundary of text block $i$
which is consistent with the pattern instance starting 
at text position $l$ (and therefore consistent with the
pattern instant starting at text position $k$ as well);
this is done using the minimum finding algorithm and takes $O(\sqrt{m})$ time
with constant failure probability.
Similarly, $h(i)$ finds the longest substring (with length at most $m/2$)
ending at the right boundary of text block $i$
which is consistent with the pattern instance starting 
at text position $k$.

Finally, using these two substrings, $h(i)$ can
determine in $O(1)$ time, whether there exists a pattern instance 
with left endpoint in block $i$ which matches the text.
If the length of the two substrings is less than $m$ then there is
no such pattern instance;
otherwise, all instances of the pattern which 
occur completely  within these two substrings and starting at shifts
of integer multiples of $|v|$ from $k$
are complete matches (here $|v|$ is  the pattern period).

Thus, $h(i)$ determines whether or not the pattern occurs in 
block $i$ in $O(\sqrt{m}\log m)$ time, with failure probability a constant.
This failure probability can be made arbitrarily close to 0
by repetition. 
Note that $h(i)$ can determine the leftmost pattern occurrence
in block $i$ as well,  if required,  within the same time bounds.

The rest of the algorithm stays the same: $h(i)$ is used to
find a block containing an occurrence of the pattern and
subsequently, an occurrence of the pattern in this block
is found using the above method.

\section{Conclusions And Open Problems}

We have shown how one occurrence or the leftmost occurrence
of $p$ in $t$ can be found in $\tilde{O}(\sqrt{n}+\sqrt{m})$ time,
with constant two-sided failure probability.
We also note that an approximate count of the number of
occurrences (within a multiplicative constant factor) can also be 
determined in $\tilde{O}(\sqrt{n}+\sqrt{m})$ using the
approximate counting algorithm of Brassard, Hoyer and Tapp
\cite{BHT}, adapted appropriately (the oracle $h(i)$
must now return a count of the number of matches rather
than just the indication of a match). 
Finally, using the same algorithm, the total number
of  occurrences of $p$ can be determined in
$\tilde{O}(\sqrt{nt}+\sqrt{m})$ time, where $t$ is the number of 
occurrences.

One open problem would be whether string matching
with don't cares can be performed in the same time bounds
as above. The main challenge here to implement convolution using 
Fast Fourier Transforms in ${\tilde O}(\sqrt{n})$ time.  
It is not obvious how this can be accomplished.


\begin{thebibliography}{99999}

\bibitem[BBHT96]{Bo}
M. Boyer, G. Brassard, P. Hoyer, A. Tapp.
Tight bounds on quantum searching.
Proceedings of {\em 4th Workshop on Physics and Computation-PhysComp},
1996,
pp. 36--43.


\bibitem[BHT]{BHT}
G. Brassard, P. Hoyer, A. Tapp.
Quantum Counting.
Proceedings of {\em 25th International Colloquium on Automata, Languages and Programming}, 1998, pp. 820--831.



\bibitem[Gr]{G}
L. Grover.
A fast quantum mechanical algorithm for database search.
Proceedings of {\em 28th ACM Symposium on Theory of Computing},
1996, pp. 212--219.

\bibitem[DH]{Du}
C. D\"urr, P. Hoyer,
A quantum algorithm for finding the minimum.
{\em Quantum Physics E-Print Archive}, 
{\tt http://xxx.lanl.gov/quant-ph/9607014}, 1996.

\bibitem[Vi]{Vi}
U. Vishkin.
Deterministic Sampling: A new technique for fast pattern matching.
{\em SIAM Journal of Computing}, 20, 1991, pp. 22--40.


\end{thebibliography}
\end{document}